\documentclass[review]{elsarticle}

\usepackage{lineno,hyperref}
\usepackage{amsmath}
\usepackage{adjustbox}
\usepackage{color}
\usepackage{multirow}
\modulolinenumbers[5]

\journal{Neurocomputing}

\begin{document}

\begin{frontmatter}

\title{On combining acoustic and modulation spectrograms in an attention LSTM-based system for speech intelligibility level classification}  





\author{Ascensi\'on Gallardo-Antol\'in}
\cortext[mycorrespondingauthor]{Corresponding author}
\ead{gallardo@ing.uc3m.es}
\address{Dept. of Signal Theory and Communications, Universidad Carlos III de Madrid\\
	Avda. de la Universidad, 30, 28911 Legan\'es, Madrid, Spain}

\author{Juan M. Montero}
\ead{juanmanuel.montero@upm.es}
\address{Speech Technology Group, ETSIT, Universidad Polit\'ecnica de Madrid\\
	Avda. de la Complutense, 30, 28040, Madrid, Spain\\
}

\begin{abstract}

Speech intelligibility can be affected by multiple factors, such as noisy environments, channel distortions or physiological issues. In this work, we deal with the problem of automatic prediction of the speech intelligibility level in this latter case. Starting from our previous work, a non-intrusive system based on LSTM networks with attention mechanism designed for this task,  we present two main contributions. In the first one, it is proposed the use of per-frame modulation spectrograms as input features, instead of compact representations derived from them that discard important temporal information. In the second one, two different strategies for the combination of \textcolor{black}{per-frame} acoustic log-mel and modulation spectrograms \textcolor{black}{into the LSTM framework} are explored: at decision level or late fusion and at utterance level or Weighted-Pooling (WP) fusion. The proposed models are evaluated with the UA-Speech database that contains dysarthric speech with different degrees of severity. On the one hand, results show that attentional LSTM networks are able to adequately modeling the modulation spectrograms sequences producing similar classification rates as in the case of log-mel spectrograms. On the other hand, both combination strategies, late and WP fusion, outperform the single-feature systems, suggesting that \textcolor{black}{per-frame} log-mel and modulation spectrograms carry complementary information \textcolor{black}{for the task of speech intelligibility prediction,} than can be effectively exploited by the LSTM-based architectures, being the system with the WP fusion strategy and Attention-Pooling the one that achieves best results.

\end{abstract}

\begin{keyword}
speech intelligibility\sep
LSTM\sep
attention\sep
acoustic spectrogram\sep
modulation spectrogram\sep
fusion
\end{keyword}

\end{frontmatter}


\section{Introduction}
\label{sec:introduction}

Speech intelligibility refers to the comprehensibility of speech and plays a crucial role in any process where oral communication is involved. Speech, and therefore its understandability level, can be demeaned due to many factors, such as background noise, reverberation or channel distortions.  Besides, certain physiological issues can give place to impairments in any of the components of the human speech production system, producing the so-called pathological or disordered voices.

In this paper, we address the problem of automatic classification of the speech intelligibility level in this latter case, and, in particular, for dysarthric voices. Dysarthria \cite{Doyle1997} is a speech disorder produced by the motor malfunctioning of the speech organs and characterized by the presence of poor phoneme articulation, imprecise transitions between adjacent phonemes, hypernasality, vocal roughness, perturbations in the elocution rate, volume and pitch, broken speech, etc. It has its origin in neurological injuries due to brain tumors, thrombotic/embolic strokes or degenerative illnesses such as Parkinson's Disease (PD). Dysarthria can serioulsy hamper the communication for patients, which can even lead to psychological problems.



\textcolor{black}{In recent years, the problem of automatic prediction of disordered speech comprehensibility has attracted the attention of numerous researchers. These studies} can be categorized into two main groups \cite{Janbakhshi2019}: \textit{intrusive} or \textit{non-blind} methods, and \textit{non-intrusive} or \textit{blind} approaches. Intrusive techniques are based on the comparison of the utterance to be evaluated to a reference model that represents non-pathological (intelligible) speech. This model is built from healthy data by using techniques such as Gaussian Mixture Models \cite{Bocklet2011} or iVectors \cite{Martinez2015}. Other works assume that the behaviour of an Speech-To-Text system trained with healthy speech when recognizing disordered utterances can be indicative of the speaker's intelligibility level \cite{Zlotnik2015}. In this sense, features derived from the recognizer output, such as the word error rate, have been proposed for this task. The main drawback of this kind of techniques is that large amounts of non-pathological data are required, although some works have recently tackled this issue \cite{Janbakhshi2019}.

In contrast to these techniques, non-intrusive methods do not rely on reference speech signals. They usually treat the automatic estimation of speech intelligibility as a regression or classification problem where the whole utterance is assigned to a single label. This kind of systems typically consist of a front-end where a set of acoustic features are computed and a back-end where the classification itself is performed. In this work, we have followed this approach.

Most of the non-intrusive systems described in the literature rely on traditional machine learning methods, such as Linear Discriminant Analysis (LDA)  \cite{Liss2010, Falk2012, SarriaPaja2012}, Support Vector Machines (SVM) \cite{Khan2014, FernandezDiaz2020} or Random Forests \cite{Byeon2018}. As these techniques do not allow the adequate modeling of temporal signals as speech, it is necessary to compute an ad-hoc \emph{utterance-level representation} at the front-end, that somehow summarizes the information contained in the corresponding temporal sequences. For example, as per-frame acoustic log-mel spectrograms\footnote{Throughout this paper, we use indistinctly the terms ``acoustic log-mel spectrogram'', ``log-mel spectrogram'' and ``log-mels''.} reflect the short-term artifacts of pathological speech, several compact features derived from them are commonly used for intelligibility prediction, as the average of mel-frequency delta-energy coefficients \cite{Hummel2011} or the average of Mel Frequency Cepstrum Coefficients (MFCC) \cite{FernandezDiaz2020}. In the same way, per-frame modulation spectrograms that convey information about long-term temporal dynamics perturbations of disordered speech are not directly used as input features. Instead, it is utilized some utterance-level characteristics derived from them, such as the frequency and amplitude of the modulation spectrum peak \cite{Liss2010}, the Low-to-High Modulation Ratio (LHMR)  \cite{Falk2012, SarriaPaja2012}  or the average energy of the modulation spectrogram \cite{FernandezDiaz2020}. Obviously, these summarized parameterizations imply an important loss of information with respect to the corresponding temporal representations, degrading the performance of the whole system.

Recently, the use of  \emph{Long Short-Term Memory} (LSTM) networks \cite{Hochreiter1997,Gers2003},  belonging to the area of Deep Learning (DL) methods, is spreading among the audio and speech research community as they are more suitable for the modeling of temporal sequences. In fact, LSTM-based systems have achieved substantial improvements in several audio and speech-related tasks, such as Acoustic Event Detection (AED) \cite{Kao2020}, Acoustic Scene Classification (ASC) \cite{Guo2017}, Speech Emotion Recognition (SER) \cite{Huang2016,Mirsamadi2017} or Cognitive Load (CL) classification from speech \cite{Gallardo-Antolin2019a, Gallardo-Antolin2019b}. In many of these works, LSTMs are combined with Weighted Pooling (WP) schemes in order to obtain an utterance-level representation that is subsequently processed by the succeeding dense layers that compose the classifier itself. Nowadays, one of the most successful WP methods is the so-called  \emph{attention pooling}  \cite{Chorowski2015,Huang2016},  a mechanism that tries to determine the structure of the temporal sequences by learning the relevance of each frame to the task under consideration.  These attentionl LSTM models have been successfully proposed for the aforementioned speech and audio tasks. A detailed review about this topic can be found in \cite{Zacarias-Morales2021}.

In our previous paper  \cite{FernandezDiaz2020}, we showed that is feasible to model per-frame acoustic log-mel spectrograms with attentional LSTM networks for the task of predicting the speech intelligibility level (low, medium or high) of a dysarthric person and we achieved significant improvements with respect to a SVM-based system with different types of ad-hoc compact acoustic features.

One of the weakness of our previous research is that it was focused on short-term perturbations of pathological speech, whereas long-term disturbances were not taken into consideration. For overcoming this problem, in this paper we extend the work into two directions. Firstly, we propose a similar attentional LSTM architecture for modeling the long-term dynamics of disordered speech. For this purpose, we consider the adoption of the per-frame modulation spectrogram as input features to the network, what, to the best of our knowledge, has not been previously reported for this task. Secondly, as it is reasonable to hypothesize that log-mel and modulation spectrograms contain complementary information for the estimation of a subject's speech intelligibility, we explore two different alternatives for the fusion of both kind of features and show that the combination improves the performance of the individual systems. \textcolor{black}{Again, to the best of our knowledge, the combination of this kind of per-frame features into a LSTM-based architecture for speech intelligibility prediction has not been previously explored in the literature.}

The rest of the document is organized as follows: Section \ref{sec:indicators} describes the motivation behind the use of acoustic log-mel and modulation spectrograms as features for the automatic estimation of the intelligibility level; Section \ref{sec:lstm}  contains the general description of the proposed attention LSTM-based intelligibility classification system, including the single-feature architecture that uses either log-mel or modulation spectrograms as inputs, and the architecture combining both kind of features; the database, experiments and results are described in Section \ref{sec:experiments}; and Section \ref{sec:conclusions} finishes the document with some conclusions and future work.

\section{Acoustic log-mel and modulation spectrograms as indicators of speech intelligibility}
\label{sec:indicators}

Several phenomena, such as vocal harshness, nasal voice, adventitious sounds, unclear transitions between adjacent phonemes, excessive phoneme duration, variable speech rate, disfluencies, presence of odd pauses, and other prosodic disturbances (pitch breaks, monotonicity, etc.), are usually present to some extent in dysarthric speech, affecting negatively its intelligibility level  \cite{DeBodt2002, Falk2012}.

Some of these artifacts can be observed by analyzing the speech recordings at short-time scales. In fact, the short-time \emph{acoustic spectrogram}, which shows the variation with time of the speech spectrum computed at short \emph{acoustic frames}, provides useful cues about the intelligibility degree of an utterance \cite{Paliwal2011}. \textcolor{black}{The acoustic spectrogram of a speech signal $x(n)$ is computed by performing a short-time analysis using the following Equation \cite{Paliwal2011},}

\textcolor{black}{
\begin{equation} 
X_a(n_a, k'_a) = \sum_{l = -\infty}^{\infty}x(l) w_a(n_a-l) e^{-j \frac{2 \pi k'_a}{N_a} }
\label{eq:acoustic_spectrogram}
\end{equation}
}

\textcolor{black}{where the subscript $a$ denotes the acoustic domain, $n_a$ is the frame index and it is related to the discrete-time index $n$ through the expression $n_a = \frac{n}{L_a}$, being $L_a$ the frame period (in samples); $k'_a$ is the index of the acoustic frequency bin; and $w_a(n) $ is the analysis window whose length is $N_a$ (in samples).}

Prior studies have concluded that the energies of certain frequency subbands suffer variations due to the presence of distorted phonemes and that those subbands placed in the limits of the  F1-F2 space are more affected to vowels' pronunciation problems \cite{Liu2005, Hummel2011}. In addition, the amount of energy above $3500~Hz - 4000~Hz$ in dysarthric speech is usually less than in non-pathological speech \cite{Kent2001, Hummel2011}. 

\begin{figure}[t]
	\centering
	\includegraphics[width=0.6\textwidth]{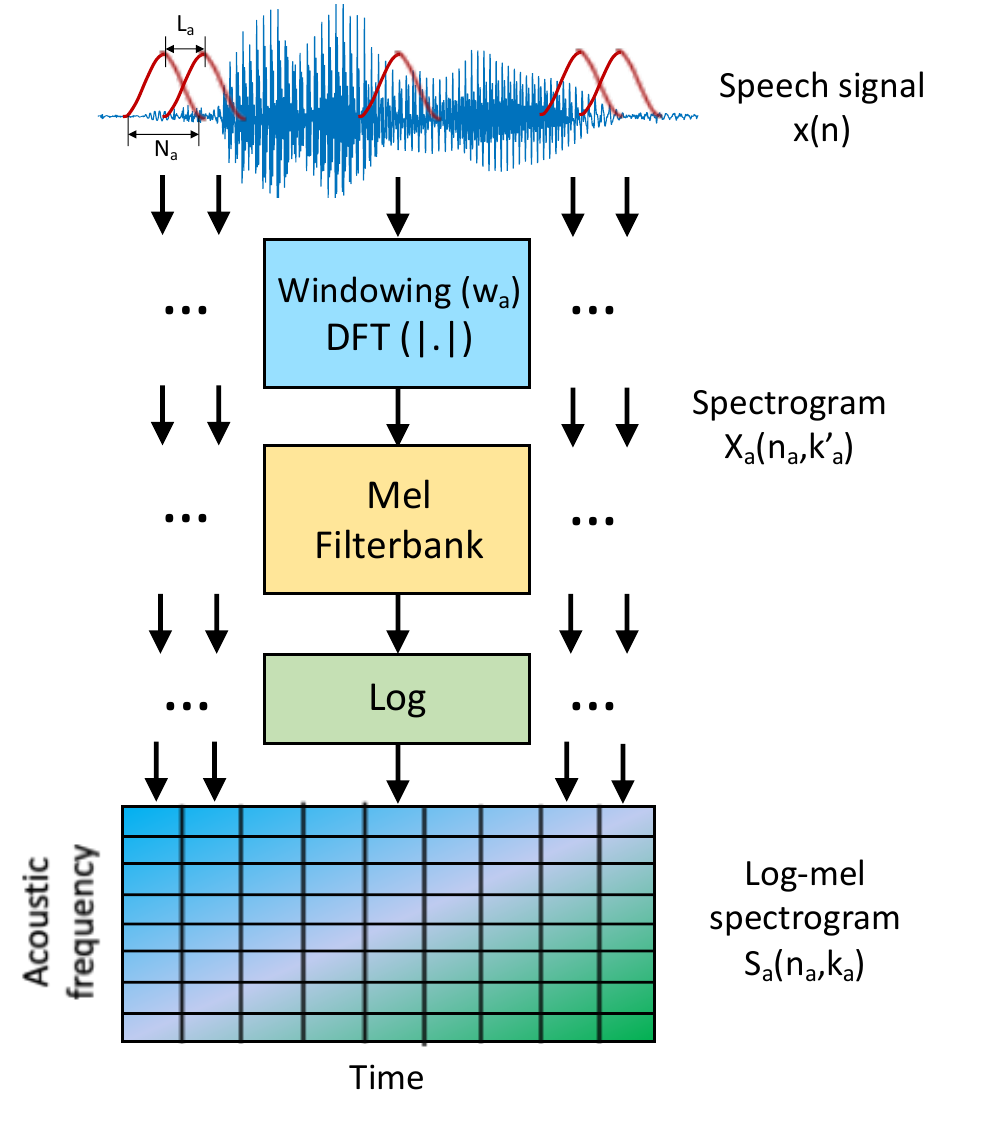}
	\caption{\textcolor{black}{Block diagram of the log-mel spectrogram computation process.}}
	\label{fig:logmel_spectrogram}
\end{figure}

Log-mel spectrograms are derived from the corresponding acoustic spectrograms that are first mapped to the mel-frequency spacing \cite{Mermelstein1976} by using an auditory filter bank composed of mel-scaled filters, and later converted to a logarithmic scale. \textcolor{black}{Log-mel spectrograms are denoted by $S_a(n_a, k_a)$, where $k_a$ is the index of the mel-scale filter and their computation process is represented in Figure \ref{fig:logmel_spectrogram}.}  One of their advantages is that they are bio-inspired spectro-temporal representations of speech, as the mel scale is a frequency warping that tries to mimic the non-equal sensitivity of the human hearing at different frequencies. As previously mentioned, when using traditional machine learning algorithms, log-mel spectrograms are not directly used as features. Instead, a compact representation, such the average of mel-frequency delta-energy coefficients \cite{Hummel2011}, the standard deviation of the first derivative of the zeroth order MFCC \cite{Falk2012, SarriaPaja2012} or the average of MFCC \cite{FernandezDiaz2020} are utilized, what could bring a relevant loss of information.

For overcoming this issue and following our previous work  \cite{FernandezDiaz2020} where it was proven that per-frame log-mel spectrograms are suitable for the task of intelligibility prediction in combination with LSTM-based classifiers, in this paper, we have chosen to use this kind of features for coping with the short-term characteristics of pathological speech. \textcolor{black}{In particular, for the log-mel spectrograms computation, we have used a Hamming window with length $N_a = 20~ms$, a frame period of $L_a = 10~ms$ and a mel-scale filterbank composed of $32$ filters.}

Dysarthric speech characteristics related to rhythmic disturbances are better reflected in representations that capture long-term speech temporal dynamics information. In particular, features derived from the \emph{modulation spectrogram}  of speech  \cite{Greenberg1997, VicentePena2006}, which captures the speed of fluctuation of long-term speech temporal envelopes,  are good candidates for this purpose. This is because it has been shown that  slow temporal envelope modulations contain information about several perturbations that can be present in pathological speech, such as, non-habitual intensity and speed variations, imprecise coarticulations or interruptions and disfluencies, and, therefore they can be used as indicators of speech intelligibility \cite{Falk2010, Paliwal2011, Falk2012}. 

From the analysis of the speech modulation spectrogram properties, previous studies have shown that in healthy voice most of the energy is located in the range of modulation frequencies from $2-20~Hz$, reaching maximum values at around $4~Hz$, which is the modulation frequency corresponding to the average syllabic rate \cite{Greenberg1997}. However, in dysarthric voice  the peak of the modulation energy suffers a shift to lower values. In addition, as the intelligibility degree decreases, the bulk of modulation energy tends to concentrate in lower modulation frequencies \cite{Liss2010, Falk2012}.

\begin{figure}[t]
 \centering
 \includegraphics[width=1.0\textwidth]{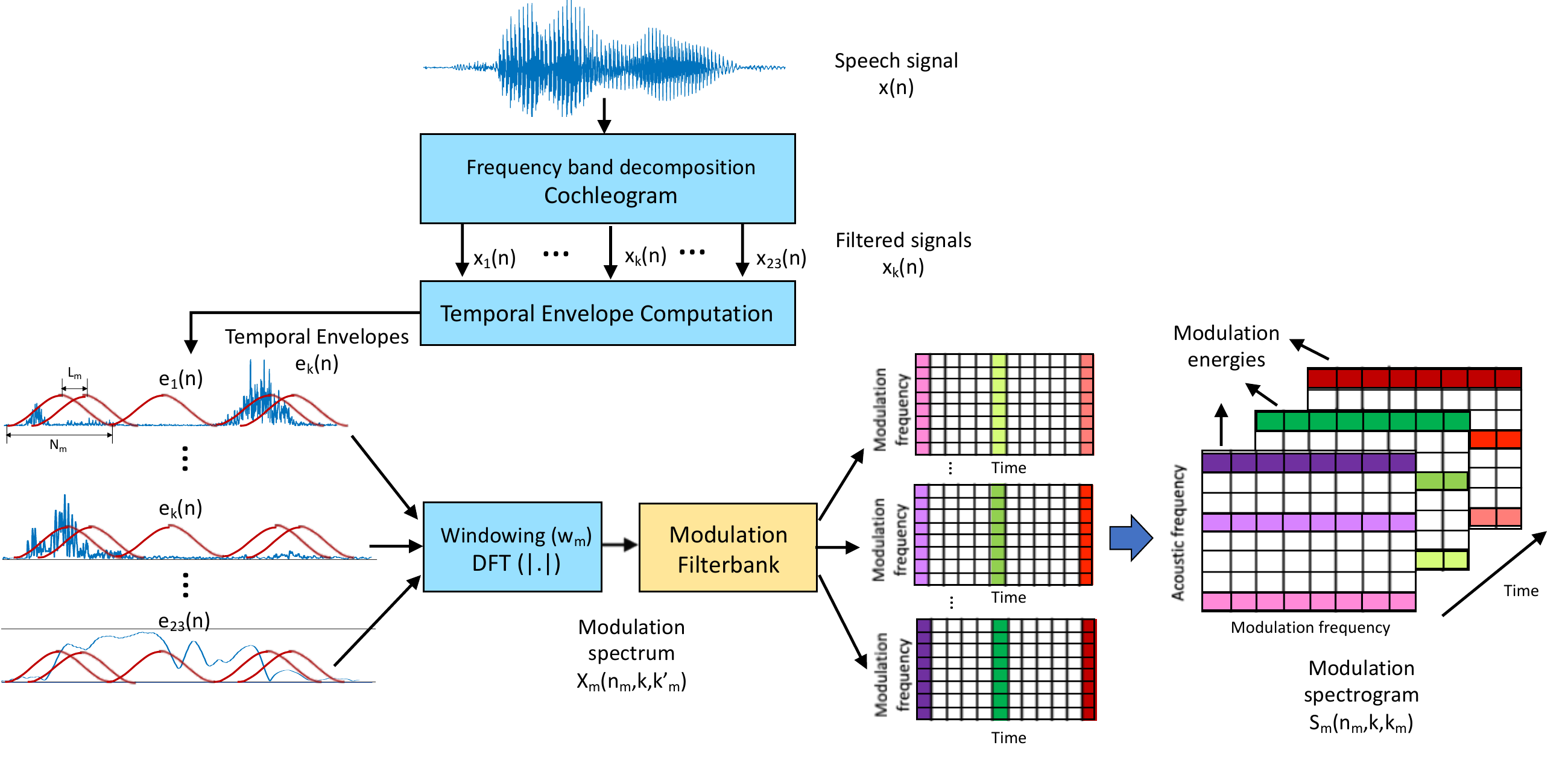}
 \caption{\textcolor{black}{Block diagram of the modulation spectrogram computation process.}}
 \label{fig:modulation_spectrogram}
 \end{figure}

In order to illustrate these previous findings, we have analyzed the modulation spectrograms of the utterances contained in the database used in our experiments (see Subsection \ref{subsec:database}). \textcolor{black}{The modulation spectrograms are obtained following the method proposed in \cite{Falk2010, SarriaPaja2017} that is summarized in Figure \ref{fig:modulation_spectrogram}. First, the speech signal $x(n)$ is decomposed into acoustic frequency bands by means of a filterbank composed of $23$  gammatone critical-band filters, resulting in a set of $23$ filtered signals $x_k(n), k = 1, ...., 23$. This filterbank mimics the cochlear processing that takes place in the human auditory system and its first and last filter are centered at $125~Hz$ and $8000~Hz$, respectively.  Then, the temporal envelopes corresponding to each frequency band, $e_k(n)$, are computed by using the Hilbert transform $H\{.\}$, as follows,}

\textcolor{black}{
\begin{equation} 
e_k(n) = \sqrt{x_k(n)^2 + H\{{x_k(n)^2}\}}
\label{eq:temporal_envelope}
\end{equation}
}

\textcolor{black}{Then, the modulation spectrum for a given acoustic frequency $k$ is computed as the Discrete Fourier Transform of the corresponding temporal envelope $e_k(n)$, as indicated in the following equation,}

\textcolor{black}{
\begin{equation} 
X_m(n_m, k, k'_m) = \sum_{l = -\infty}^{\infty}e_k(l) w_m(n_m-l) e^{-j \frac{2 \pi k'_m}{N_m} }
\label{eq:modulation_spectrogram}
\end{equation}
}

\textcolor{black}{where the subscript $m$ denotes the modulation domain, $n_m$ is the frame index and it is related to the discrete-time index $n$ through the expression $n_m = \frac{n}{L_m}$, being $L_m$ the frame period (in samples); $k'_m$ is the index of the modulation frequency bin; and $w_m(n) $ is the analysis window whose length is $N_m$ (in samples). In this work, $w_m(n)$ is a  Hamming window with length $N_m = 256~ms$ and the frame shift is $L_m = 64~ms$. The modulation frequencies $k'_m$  are grouped according to a modulation filterbank composed of $8$ second-order bandpass filters with a quality factor $Q = 2$, whose center frequencies are located in the range $2-64~Hz$, yielding the modulation spectrogram that is denoted as $S_m(n_m, k, k_m)$, where $k_m$ is the index of the modulation filter. Note that for each \emph{modulation frame} $n_m$, it is obtained a 2D representation that contains the modulation energies, whose dimensions are the number of acoustic frequencies multiplied by the number of modulation frequencies (in this case, $23 \times 8$).}
 
\begin{figure}[t]
 	\centering
 	\includegraphics[width=1.0\textwidth]{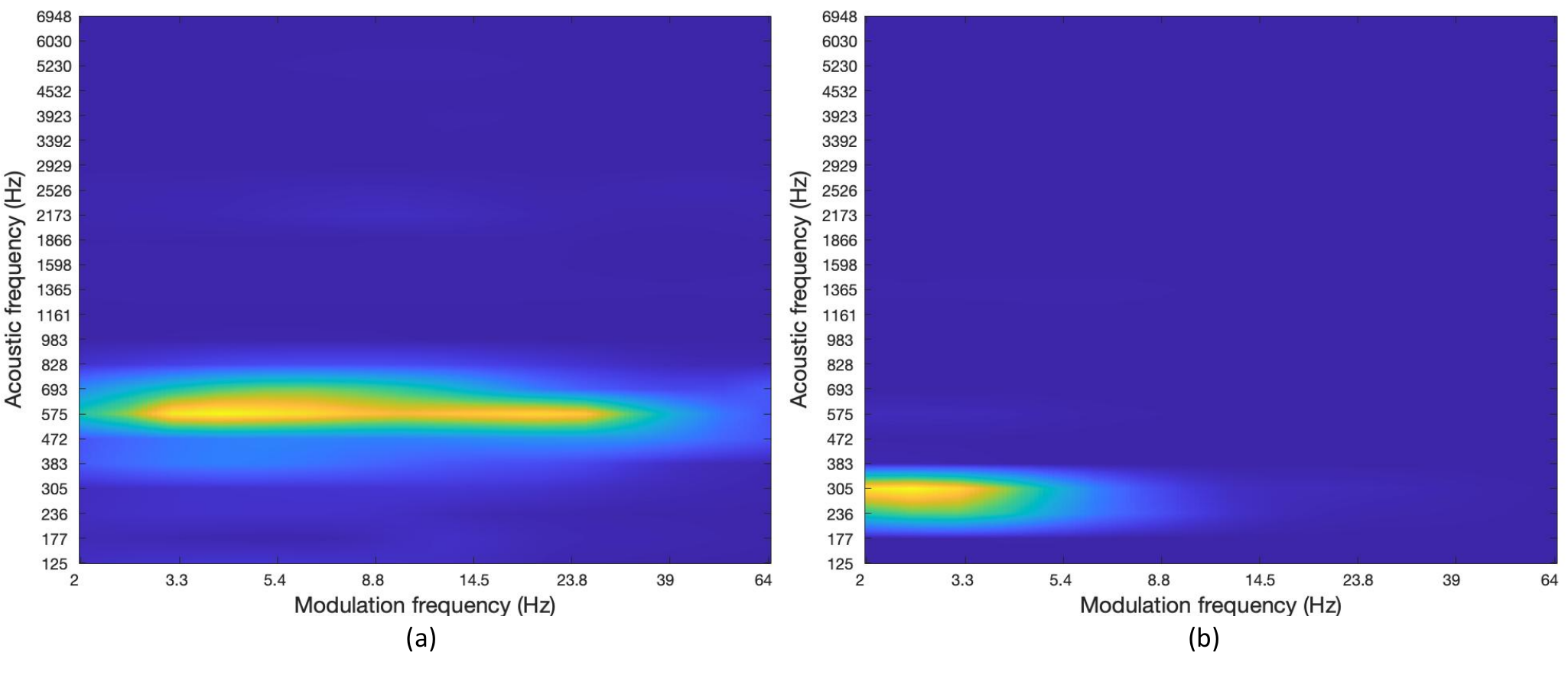}
 	\caption{Modulation spectrum of a certain modulation frame belonging to a speech recording with (a) high intelligibility and (b) low intelligibility. Both frames correspond to the central part of the first vowel of the word ``jowls".}
 	\label{fig:ms_frame}
 \end{figure}

Figure \ref{fig:ms_frame} shows two examples of the modulation spectrum of a particular modulation frame belonging to an utterance with high (a) and low intelligibility (b). In both cases, the horizontal and vertical axes represent, respectively, the modulation and acoustic frequencies. It can be observed that for the low intelligibility speaker, the modulation energies are concentrated in low modulation frequencies, being the maximum values located below $4~Hz$ (see Figure \ref{fig:ms_frame} (b)), whereas in the case of the high intelligibility speaker, the energy peak is situated around $4~Hz$ and the modulation energy spreads over higher modulation frequencies (see Figure \ref{fig:ms_frame}(a)).
 
\begin{figure}[t]
 	\centering
 	\includegraphics[width=1.0\textwidth]{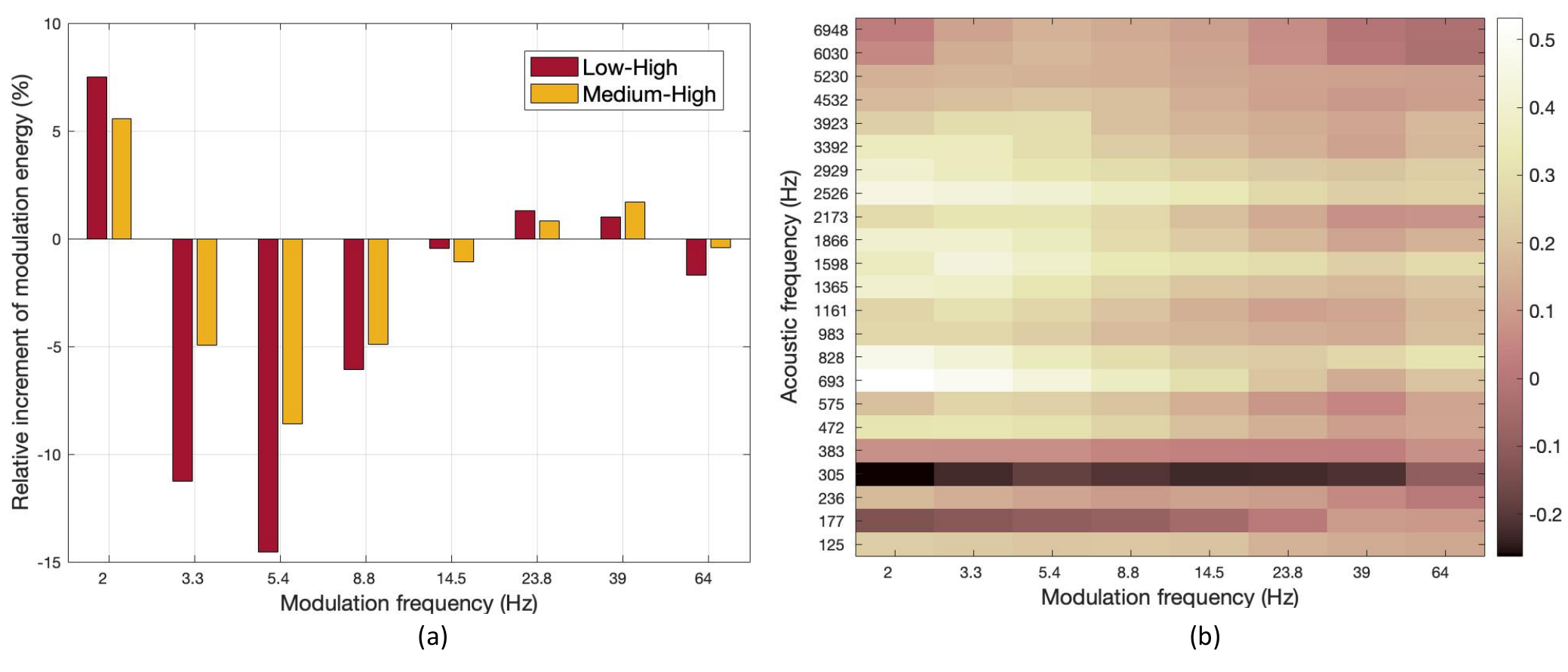}
 	\caption{Relationship between modulation energy and speech intelligibility level: (a) Average relative increment/decrement of energy per modulation band for low and medium intelligibility utterances with respect to the energy of the corresponding band of the high intelligibility recordings; (b) Correlation values between modulation energy and intelligibility level.}
 	\label{fig:incr_corr_energy}
 \end{figure}

 Figure \ref{fig:incr_corr_energy}(a) shows the average relative increment/decrement of energy per modulation band for low and medium intelligibility utterances with respect to the energy of the corresponding band of the high intelligibility recordings. These values were computed from one of the folds of the dataset used in our experimentation (see Subsection \ref{subsec:database}). Energies per modulation band were computed by averaging, for each modulation band, the energies across the different acoustic frequencies and normalizing them by the total modulation energy of the utterance. As this figure shows, in the case of low intelligibility, low-frequency modulation energy is slightly increased with respect to high intelligibility speech, whereas in the range of approximately $3~Hz - 10~Hz$, the opposite occurs.  This fact corroborates the observations drawn in the aforementioned studies \cite{Liss2010, Falk2012}. For medium intelligibility, same trends are observed although the increment/decrement values are smaller than in the previous case. For all intelligibility levels, the absolute energy located in very high-frequency regions is small, so its relative increments or decrements do not seem to provide a significant information for our task.   
 
Figure \ref{fig:incr_corr_energy}(b) represents the correlation values between the per-frame modulation energy and the intelligibility level. These values have been calculated from the same fold of the dataset than in the previous case. On the one hand, it can be seen that the region that is directly correlated with the intelligibility degree is roughly the one with a modulation frequency below $10~Hz$ and an acoustic frequency from $450~Hz$ to $3500~Hz$. On the other hand, the intelligibility level seems to be slightly inversely correlated with energies at acoustic frequencies around $300~Hz$ independently of the modulation frequency. Although, in general, the individual correlations are not very high, it seems apparent that the modulation spectrogram at a whole contains useful cues for the inference of the intelligibility, what justifies the use of automatic learning techniques able of exploiting this information. 
 
In previous research, it has been proposed several features derived from the modulation spectrogram for speech intelligibility prediction in combination with traditional machine learning classifiers, as LDA \cite{Liss2010, Falk2012, SarriaPaja2012} or SVMs  \cite{Khan2014, FernandezDiaz2020}. The Low-to-High Modulation energy Ratio (LHMR), which is the quotient of modulation energy at modulation frequencies below $4~Hz$ to modulation frequencies above $4~Hz$ \cite{Liss2010, Falk2012}, frequency and amplitude of the modulation spectrum peak, energy in the region of $3–6~Hz$ \cite{Liss2010} or the average of modulation energies across the frames \cite{FernandezDiaz2020} are good examples. However, as these features can be seen as summarizations of the modulation spectrogram content, their use likely implies a loss of temporal information.

Regarding this issue, our conjecture is that, as in the case of log-mel spectrograms, DL-based models are able to exploit all the information conveyed in the per-frame modulation spectrograms, allowing their effective use in an automatic intelligibility classification system. In particular, due to the temporal nature of the per-frame modulation spectrograms, we propose the use of LSTMs for this purpose, as explained in Section \ref{sec:lstm}.

Besides, recalling that dysarthric speech characteristics are related to short and long-term phenomena, we argue that a combination of features extracted at different time scales is required in order to better determine the intelligibility level of an utterance. For this reason, we propose the fusion of log-mel and modulation spectrograms in the framework of a LSTM-based system as described in Subsection \ref{subsec:combined_systems}.

\section{LSTM-based speech intelligibility classification system}
\label{sec:lstm}

Long Short-Term Memory networks are a special type of Recurrent Neural Networks that are able to learn long-term dependencies due to their capacity to store past information in their memory blocks \cite{Hochreiter1997,Gers2003}, in such a way that their outputs depend on past and present  inputs. For that, LSTMs  are very suitable for modeling temporal sequences, such as  log-mel and modulation spectrograms that are the feature inputs considered in this work.

LSTMs perform a  sequence-to-sequence learning where an input sequence of length $T$, $x = \{x_1, ..., x_T\}$ is converted to 
an output sequence $y = \{y_1, ..., y_T\}$ of the same length. As in the speech intelligibility classification problem, a single label (intelligibility level) must be assigned to the whole input sequence, a Weighted Pooling (WP) stage is connected to the LSTM layer, whose purpose is to aggregate the information contained in $y$ and produce an utterance-level representation $z$ that, in turn, is the input to the classifier itself \cite{Huang2016, Huang2017}.  \textcolor{black}{WP is commonly implemented as a simple weighted aggregation operation as follows,}

 \textcolor{black}{
\begin{equation}
z = \sum_{t=1}^{T} \alpha_t y_t
\label{eq:wp}
\end{equation}
}

 \textcolor{black}{where $y = \{y_1, y_2, ..., y_T\}$ is the LSTM output sequence, $\alpha = \{\alpha_1, \alpha_2, ..., \alpha_T\}$ is the weight vector and $z$ is the final utterance-level representation. The way weights are computed determines different types of WP  approaches. In this work, we have considered the following ones.}

\paragraph{Basic LSTM}
In this case, it is assumed that the last frame of the LSTM output, $y_T$, is the most representative one, as information from the whole sequence has been used to some extent for its computation \cite{Zazo2016}. Therefore, only this last frame goes into the succeeding layer, discarding the remaining ones, i.e., $z = y_T$.  \textcolor{black}{In other words, all the weights are zero except the one corresponding to the last LSTM frame that is equal to one (i. e., $\alpha_t = 0, t = 1...., T-1$ and $\alpha_T = 1$).}

\paragraph{Mean-Pooling}
In the Mean-Pooling approach, the utterance-level representation is calculated as the average of the output LSTM frames along the whole sequence, \textcolor{black}{as indicated in the following expression,}

 \textcolor{black}{
	\begin{equation}
	z = \frac{1}{T} \sum_{t=1}^{T}  y_t
	\label{eq:mean_pooling}
	\end{equation}
}

\textcolor{black}{In this case, it is assumed that all the LSTM frames are equally important. Therefore, all the weights are equal ($\alpha_t = \frac{1}{T},\,\, \forall \, t$)  and, as a consequence, all the elements of $y$ contribute evenly to $z$.}

\paragraph{Attention-Pooling}
The rationale behind this approach is that not all the LSTM frames reflect the understandability level with the same intensity. For this reason, frames containing more cues about the intelligibility degree should be more emphasized than the remaining ones. This way, the utterance-level representation $z$ is computed as the weighted arithmetic mean of the output LSTM frames, where larger weights should be set to the more relevant frames to the task, whereas smaller weights should be assigned to frames not conveying useful information.

For the computation of the weights, we have adopted the method proposed by  \cite{Mirsamadi2017}  for speech emotion recognition, that is especially suitable for scenarios where the amount of training data is limited, as is our case. In this approach, \textcolor{black}{the unnormalized attention weights, $\bar{\alpha} = \{\bar{\alpha_1}, \bar{\alpha_2}, ..., \bar{\alpha_T}\}$},  are calculated through the following Equation \eqref{eq:attention_weights}.

\textcolor{black}{
\begin{equation} 
\bar{\alpha_{t}}=\frac{\exp \left(u^{\text{tr}} y_t\right)}{\sum_{t=1}^{T} \exp \left(u^{\text{tr}} y_t\right)},
\label{eq:attention_weights}
\end{equation}
}\label{key}

where $u$ and $y$ are, respectively, the attention parameter vector and the LSTM output, and the superscript $tr$ denotes a transpose operation. The inner product between $u$ and $y_t$ measures the relevance of each $t$-th frame. In order to obtain a set of normalized weights whose sum across all the frames of the sequence is equal to one, \textcolor{black}{a softmax transformation is applied to these quantities, according to the following expression,}

\textcolor{black}{
\begin{equation}
\alpha_t = \frac{\text{exp}(\bar{\alpha_{t}}(t))}{\sum_{t=1}^{t=T}\text{exp}(\bar{\alpha_{t}}(t))}
\label{eq6}
\end{equation}
}

The attention parameters and the LSTM outputs are obtained in the training process of the network.

In the following Subsections, we describe the architectures proposed for the single-feature LSTM-based system, where using only log-mel spectrograms or modulations spectrograms as input, and the combined LSTM-based systems, where both, log-mel and modulations spectrograms are jointly utilized as input features.

\subsection{Single-feature LSTM-based architecture}
\label{subsec:individual_systems}

\begin{figure}[t]
	\centering
	\includegraphics[width=0.25\textwidth]{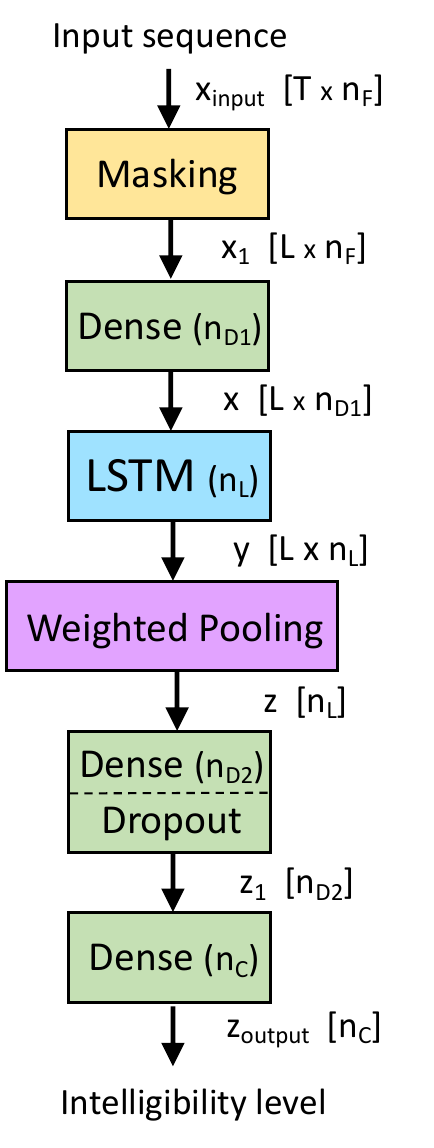}
	\caption{LSTM-based architecture for speech intelligibility classification when using either log-mel spectrograms or modulation spectrograms as input features. The dimensions of each variable are indicated in brackets.}
	\label{fig:lstm_arch_single}
\end{figure}

Figure \ref{fig:lstm_arch_single}(a) shows the block diagram of the single-feature LSTM system proposed for speech intelligibility classification, which is based on our previous work \cite{FernandezDiaz2020}. As can be seen, the general architecture of the system is the same, no matter the input consists of log-mel or modulation spectrograms. However, the specific values of  the variables in the model are different depending of the kind of input features considered, as is detailed in Section \ref{sec:experiments} and Table \ref{tab:lstm_configuration}. 

The dimension of the input features $x_{input}$ are $T \times n_F$, being $T$ the number of acoustic or modulation frames of the speech signal and $n_F$ the number of components of the feature vectors. In the case of log-mel spectrograms $n_F$ matches the number of mel filters, whereas in the case of modulation spectrograms $n_F$ is equal to the number of acoustic frequencies multiplied by the number of modulation frequencies considered in the modulation spectra computation. 

As both, log-mel and modulation spectrograms do not have the same length for all the speech recordings and, however, the length of the LSTM input sequences is set to a fixed value $L$, a preprocessing of the sequences is required in order to overcome this issue. In particular, longer sequences than  $L$ are cut (this only occurs in a few cases as it is shown in Subsection \ref{subset:feature_extraction}), whereas shorter sequences are padded with masked values. A masking layer allows that these dummy values not to be used in further computations.  Note that the value of $L$ is different for each type of feature. This is because log-mel and modulation spectrograms are extracted at different frame periods as indicated in Subsection \ref{subset:feature_extraction}.

Next, a dense layer of $n_{D1}$ neurons is connected to the masking layer. Its purpose is to perform some kind of feature extraction for improving the input to the following LSTM layer. This consists of  $n_L$ cells, being the output of each cell a sequence of length $L$. The information of these LSTM sequences is compressed by using a Weighted Pooling stage. As aforementioned, three strategies for WP have been considered: basic LSTM, Mean-Pooling and Attention-Pooling.

The WP module is connected to a second dense layer of $n_{D2}$ neurons with dropout in order to avoid overfitting. In the end, its output goes into a final dense layer with $n_C$ nodes that is activated by a softmax function for performing the multiclass classification. The value of $n_C$ matches the number of possible intelligibility levels to be predicted (low, medium and high).

\subsection{Fusion strategies for the combination of log-mel and modulation spectrograms}
\label{subsec:combined_systems}

\begin{figure}[t]
	\centering
	\includegraphics[width=1.0\textwidth]{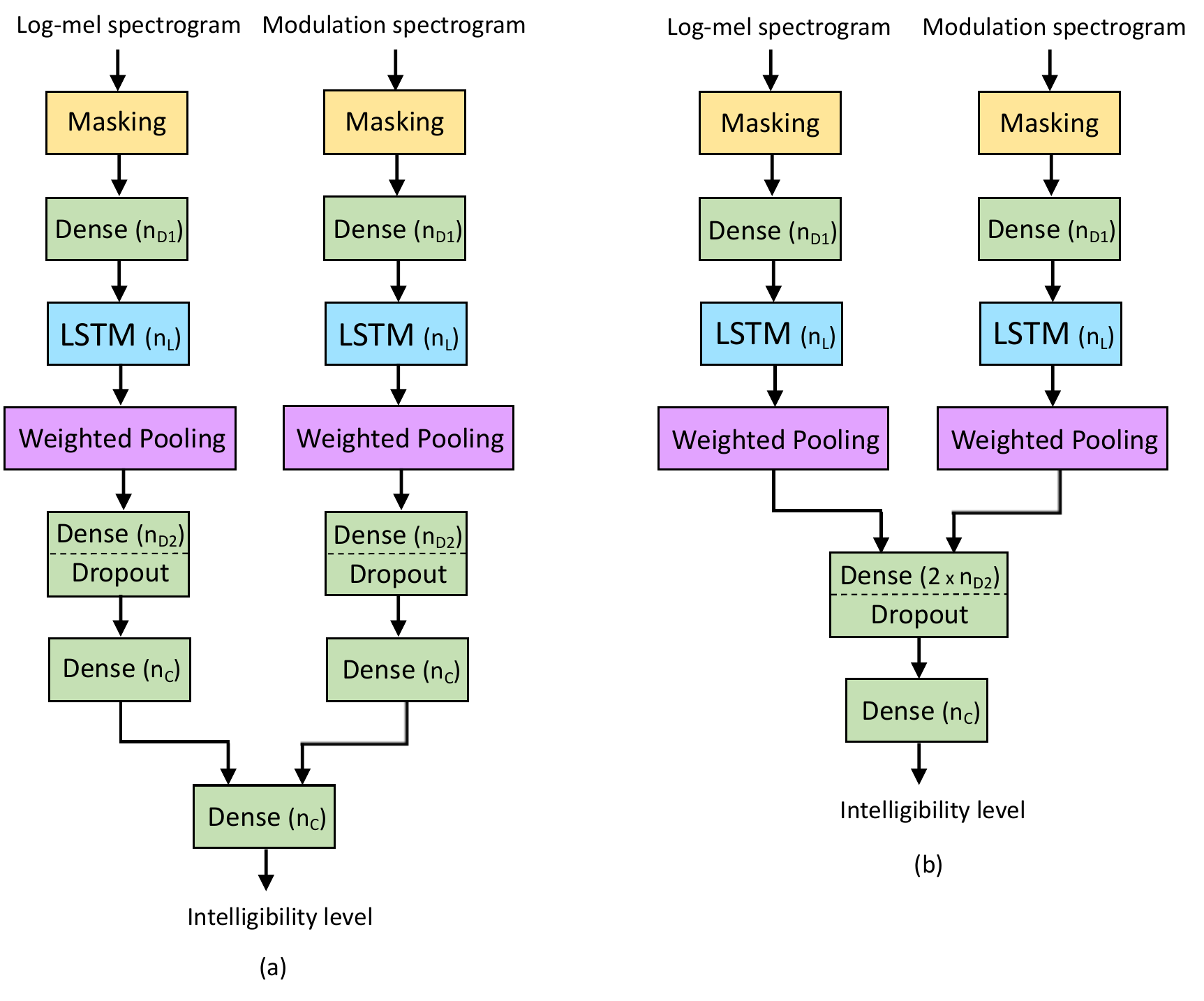}
	\caption{Fusion strategies for the combination of log-mel and modulation spectrograms. $n_{D1}$, $n_L$, $n_{D2}$ and $n_C$ stand for the number of neurons in the first dense layer, the number of LSTM units, the number of neurons in the second dense layer of the individual systems and the number of classes (intelligibility levels), respectively. (a) Late fusion: combination at decision level; (b)  WP fusion: combination at utterance level.}
	\label{fig:lstm_arch_comb}
\end{figure}

We have explored two fusion strategies for the combination of log-mel and modulation spectrograms, namely, \emph{late fusion} and \emph{WP fusion}, as depicted in Figure \ref{fig:lstm_arch_comb}. In the first case, the combination is produced at decision level by means of a dense layer with $n_C$ neurons and sigmoid activation whose inputs are directly the final outputs of the individual systems. In the second case, the fusion is performed at utterance level, i.e., the outputs of the Weighted Pooling layers of the individual systems are combined by using a dense layer of $2 \times n_{D2}$ nodes with dropout, followed by a final fully-connected layer of $n_C$ neurons and sigmoid activation.
 
\section{Experiments}
\label{sec:experiments}

\subsection{Database}
\label{subsec:database}

For our experiments, we have used the UA-Speech database \cite{Kim2008} that consists of utterances (digits, computer commands, simple short words, complex long words and the radio alphabet) pronounced by 15 persons (11 men and 4 women) suffering from dysarthria with different degrees. All the speech files were recorded  at $16~KHz$ with an array of 7 microphones, although only signals corresponding to the sixth microphone have been utilized. The dataset also contains speech from healthy control subjects. However, these audio recordings have been discarded, as in this work we have followed the non-intrusive approach for the design of the speech intelligibility classification system. In summary, the total number of available files for the experiments is $11,435$.

The database was manually annotated by medical staff  in terms of the intelligibility score, that can be defined as the average percentage of understood words by the specialists after carrying out a series of subjective tests. As intelligibility scores range from 0 (completely unintelligible) to 100 (perfectly intelligible), these original labels were mapped to the three categories considered in this work. This way, the low, medium and high intelligibility classes correspond to, respectively, scores from 0 to 33, from 34 to 66 and from 67 to 100.

Regarding the experimental protocol, we have used a subject-wise 5-fold cross validation. Specifically, the database was split into five disjoint balanced groups, in such a way that all recordings from the same subject were included in the same subset. In each fold,  one group was kept for testing, another different group was utilized for validation, whereas the remainder ones were used for training. The experiments were repeated five times alternating the training, validation and test sets, averaging the  results afterwards. We adopt this speaker-independent configuration in order to prevent the system to learn the speaker's identity instead of his/her intelligibility level.

\subsection{Feature extraction}
\label{subset:feature_extraction}

Log-mel spectrograms were computed using a Hamming window of duration \textcolor{black}{$N_a = 20~ms$  with a frame shift of $L_a = 10~ms$} and a filterbank composed of $32$ triangular filters distributed according to the mel scale. For this purpose, the  Python's package LibROSA \cite{LibROSA} was utilized. Mean and standard deviation normalization were applied at utterance-level obtaining a set of normalized log-mel spectrogram sequences $x_{input}$ with $T \times n_F$ dimensions, where $T$ is the number of acoustic frames of each utterance (note that each acoustic frame corresponds to $10~ms$) and $n_F$ is the number of mel filters, i.e., $n_F =  32$.

Modulation spectrograms were computed over windows of \textcolor{black}{duration $N_m = 256~ms$ with a frame shift of $L_m = 64~ms$} using $23$ critical band frequencies and $8$ modulation filters according to the recommendations in \cite{Falk2012, SarriaPaja2012,SarriaPaja2017}, and by using the Matlab software SRMR \cite{Falk2010}. For each modulation frame, a 2D modulation spectrum was calculated and subsequently flattened, producing a vector of  $23 \; \text{x} \; 8 = 184$ dimensions. Again, mean and standard deviation normalization were applied at utterance-level yielding a set of normalized modulation spectrogram sequences $x_{input}$ with $T \times n_F$ dimensions, being, in this case, $T$ the number of modulation frames (note that  each modulation frame corresponds to $64~ms$)  and $n_F$ the dimension of each vectorized modulation spectrum, i.e., $n_F = 184$. 

For the efficient computation of LSTM networks, it is required that all the input sequences have the same dimensions, and therefore, a fixed length was established for all audios. In particular, the maximum length was set to $7~s$  because  more than $95\%$ of the audio signals were shorter than this quantity (see \cite{FernandezDiaz2020} for details). This duration corresponds to  $L = 700$ and  $L = 110$ for,  respectively, log-mel and modulation spectrograms.  Feature sequences longer than $L$  were cut, and, otherwise, they were padded with masked values that are ignored in further computations by means of the use of the appropriate masking layers in the LSTM-based architectures.

\subsection{LSTM-based classifiers}
\label{subset:classifiers}

All the individual LSTM-based systems (Basic, Mean-Pooling and Attention-Pooling) for the two kind of features (log-mel and modulation spectrograms), as well as the corresponding fused systems, were implemented with the Python's packages Tensorflow \cite{Abadi2015} and Keras \cite{Chollet2015}. The specific values of the configuration parameters of the single and combined architectures depicted, respectively,  in Figures \ref{fig:lstm_arch_single} and \ref{fig:lstm_arch_comb} are detailed in Table \ref{tab:lstm_configuration}. In all cases, the LSTM models were trained using stochastic gradient descent and the Adam optimization algorithm with an initial learning rate of $0.0002$,  a batch size of 32 and a maximum number of 50 epochs. In order to avoid overfitting in the training process, a $33 \%$ dropout was set in certain layers as indicated in Figures \ref{fig:lstm_arch_single} and \ref{fig:lstm_arch_comb}.

In the Attention-Pooling systems, the attention parameter vector $u$ had a dimension of $n_L = 64$ and its components were initialized to $1/L$, i.e., to $1/700$ and $1/110$ for log-mel and modulation spectrograms, respectively.

\begin{table}[t]
	\begin{center}
		\begin{adjustbox}{max width=1.0\textwidth}
			\begin{tabular}{|c|c|c|c|}
				\hline 
				\multirow{3}{*}{Parameter}  & \multirow{3}{*}{Description} & \multicolumn{2}{ |c|}{Value}  \\
				\cline{3-4}
				&  & Log-mel & Modulation \\
				&  & spectrogram & spectrogram \\
				\hline \hline 
				\multirow{2}{*}{$T$}  & \multirow{2}{*}{No. of frames of the input signal}  & Variable -  & Variable -  \\
				&  & Range: [121, 5691] & Range: [19, 890]   \\
				\hline
				\multirow{2}{*}{$n_F$}  & No. of features in the log-mel /  & \multirow{2}{*}{32}   & \multirow{2}{*}{184 }  \\
                & modulation spectrogram per frame  & & \\
				\hline 
				$L$ & No. of frames of the LSTM input/output sequences & 700 & 110 \\
				\hline 
				$n_{D1}$ & No. of neurons in the first dense layer & 32 & 100 \\
				\hline 
				$n_L$ & No. of LSTM units & 64 & 64 \\
				\hline
				$n_{D2}$ & No. of neurons in the second dense layer & 25 & 25  \\
				\hline 
				$n_C$ & No. of classes (intelligibility levels) & 3 & 3\\
				\hline 
			\end{tabular} 
	  \end{adjustbox}
	\end{center}
	\caption{Values of the configuration parameters for the LSTM-based systems.}
	\label{tab:lstm_configuration}
\end{table}

\subsection{\textcolor{black}{Reference systems}}
\label{subset:reference_system}

\textcolor{black}{For comparison purposes, several reference systems based on the traditional machine learning technique SVM were implemented using the MATLAB Statistics and Machine Learning toolbox.}

\textcolor{black}{Firstly, two single-feature SVM-based systems were developed considering two different types of  compact parameterizations: the average of the MFCC and the average of the modulation spectrogram. The first parameterization, MFCC, is a very popular feature extraction procedure in audio and speech related tasks (see, for example, \cite{Gallardo-Antolin2010,SarriaPaja2017}), and  for this reason, it was tried for the task under consideration in our previous work \cite{FernandezDiaz2020}. MFCCs are extracted on a frame-by-frame basis by applying the Discrete Cosine Transform on the log-mel spectrogram of the speech signal (see Subsection \ref{subset:feature_extraction}) and retaining the first $13$  coefficients. These coefficients are augmented by adding their first derivatives. Finally, the average of these parameters over all the acoustic frames are computed, yielding a final vector composed of $26$ components per utterance. The second parameterization consists of the average across all the modulation frames of the energies of the modulation spectrogram, which is obtained following the procedure mentioned in Subsection \ref{subset:feature_extraction}. In this case, each utterance is represented by a $184$-dimension vector.}

\textcolor{black}{Secondly, two more SVM-based systems were implemented for assessing the combination of the two aforementioned sets of acoustic characteristics. In the first case, systems were fused at decision level (\emph{late fusion}). In particular, the scores produced by each of the individual SVM systems were transformed to probabilities by applying a softmax operation, and then multiplied between them for obtaining the final score. In the second case, the fusion was done at feature level by means of the concatenation of the average of MFCC and the average of the modulation spectrogram into a single feature vector (\emph{early fusion}).}

\textcolor{black}{In all systems, a Radial Basis Function (RBF) kernel was used and the optimal hyperparameters of the model were obtained by means of a Bayesian optimizer with a 5-fold cross validation strategy.}

\subsection{Results}
\label{subsec:results}

The developed systems were assessed in terms of the \emph{accuracy or classification rate per audio recording}, that is defined as the percentage of correctly classified files with respect to the total number of tested files. In all cases, each experiment was run 20 times and therefore, results reported in the tables contained in this Section are given as the average accuracy across the 20 subexperiments together to the corresponding standard deviation.

\paragraph{Results of the individual systems}

Table \ref{tab:results_single} contains the accuracies achieved by the single-feature LSTM-based systems with the two different parameterizations under consideration: log-mel and modulation spectrograms, and the three Weighted Pooling schemes studied in this work: Basic, Mean-Pooling and Attention-Pooling. \textcolor{black}{For comparison purposes, classification rates obtained by the reference SVM-based systems with the average of MFCC and with the average of the modulation spectrograms as input features are also reported.}

\begin{figure}[t]
	\centering
	\includegraphics[width=1.0\textwidth]{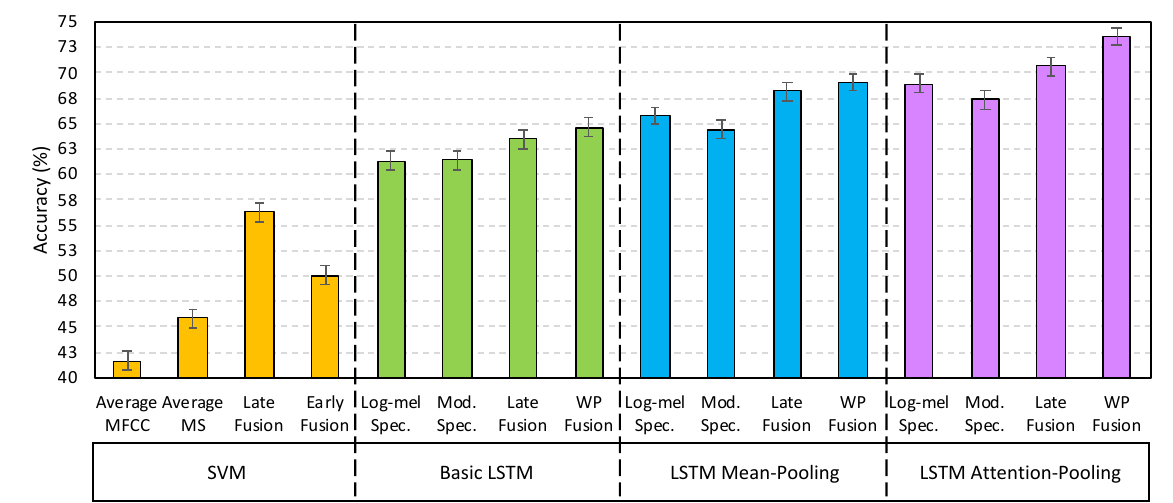}
	\caption{\textcolor{black}{Average classification rates [\%] and the corresponding  95\% confidence intervals for both, SVM-based and LSTM-based systems.}}
	\label{fig:results}
\end{figure}

\textcolor{black}{In order to analyze the statistical significance of the results, Figure \ref{fig:results} depicts the recognition rates achieved by all the systems evaluated in this work, together to the corresponding 95\% confidence intervals.}

\textcolor{black}{Firstly, it can be observed that the performance of the reference SVM-based system is poor regardless of the input features, and significantly worse than any of the LSTM-based systems. This result suggests the importance of correctly modeling the temporal behavior of the acoustic features.}

\textcolor{black}{Secondly}, the classification rates obtained \textcolor{black}{by LSTM} when using log-mel spectrograms as input features are analyzed. Note that these results are not directly comparable  to those reported in our previous paper \cite{FernandezDiaz2020}  because here we adopted a 5-fold cross validation strategy as mentioned in Subsection \ref{subsec:database}, whereas in  \cite{FernandezDiaz2020}  a fixed dataset partition ($50~\%$ for training, $15~\%$ for validation and \textcolor{black}{$35~\%$ for test})  was used. As can be observed, the simplest WP strategy, (\emph{Basic LSTM - No Pooling}), which corresponds with the case where only the last LSTM frame is considered for classification, produces the worst result. The most plausible explanation is that with this approach useful information contained in the remaining LSTM frames is ignored. In fact, results improve when the Mean-Pooling strategy (\emph{LSTM Mean-Pooling}) is applied, as, in this case, all the LSTM frames contribute with the same weight to the utterance-level representation. Nevertheless, further improvements can be achieved when incorporating the attention mechanism into the LSTM framework (\emph{LSTM Attention-Pooling}), since it allows the system to learn what are the most relevant parts of the utterances with regard to speech intelligibility estimation, diminishing the contribution of the remaining ones.  In particular, \emph{LSTM Attention-Pooling}  obtains a relative error reduction of $19.61~\%$  with respect to \emph{Basic LSTM - No Pooling} and of $9.23~\%$ with respect to \emph{LSTM Mean-Pooling}. 

\textcolor{black}{Thirdly}, as for the use of modulation spectrograms \textcolor{black}{with LSTM}, results show that this kind of features contain some important cues about the comprehensibility of an utterance than can be effectively exploited by means of LSTM-based classifiers. Regarding the performance of the different WP approaches, again, the \emph{LSTM Attention-Pooling} system clearly outperforms the other two WP strategies, achieving relative error reductions of $15.41~\%$ and $8.26~\%$ with respect to \emph{Basic LSTM - No Pooling}  and  \emph{LSTM Mean-Pooling}, respectively. This behavior suggests that it is important to emphasize the contribution of the more significant frames for the task independently of the type of features used.
	
Finally, regarding the comparison between the two kind of parameterizations \textcolor{black}{in the LSTM framework}, results with log-mel spectrograms are very similar to those achieved by modulation spectrograms for the \emph{Basic LSTM - No Pooling}  system. However, for the  \emph{LSTM Mean-Pooling} and \emph{LSTM Attention-Pooling} approaches, log-mels perform better than modulation spectrograms, although the differences are not very noticeable. \textcolor{black}{In fact, they are not statistically significant as is observed in Figure \ref{fig:results} . This result suggests that modulation spectrograms can be used as an alternative parameterization to log-mels. Moreover, in next paragraphs, it will be shown that both kind of features can be utilized in a complementary manner.} 

\begin{table}[t]
	\centering	
	
	\begin{tabular}{|c|c|c|}
		\hline
		\textbf{System} & \textbf{Features} & \textbf{Accuracy [\%]}  \\
		\hline
		\hline
		\textcolor{black}{SVM} & \textcolor{black}{Average MFCC}  & 	\textcolor{black}{41.68~\%  $\pm$  0.85~\%}  \\
		\hline
		\textcolor{black}{SVM} & \textcolor{black}{Average modulation spec.} & \textcolor{black}{45.81~\%  $\pm$  0.71~\%} \\
		\hline
		\hline 
		Basic LSTM - No Pooling & Log-mel spectrogram & 61.35~\%  $\pm$  0.28~\% \\
		\hline
		LSTM Mean-Pooling & Log-mel spectrogram & 65.77~\%  $\pm$  0.47~\% \\
		\hline
		LSTM Attention-Pooling & Log-mel spectrogram & \textbf{68.93~\%}  $\pm$  \textbf{0.35~\%} \\
		\hline
		\hline
		Basic LSTM - No Pooling & Modulation spectrogram & 61.38~\%  $\pm$  0.26~\% \\
		\hline
		LSTM Mean-Pooling & Modulation spectrogram & 64.39~\%  $\pm$  0.36~\% \\
		\hline
		LSTM Attention-Pooling & Modulation spectrogram & \textbf{67.33~}\%  $\pm$  \textbf{0.54~}\% \\
		\hline
	\end{tabular}	
	\caption{\textcolor{black}{Average classification rates [\%] and the corresponding standard deviations achieved by the SVM-based reference system with either average of MFCC or the average energy of the modulation spectrogram as input features,} and the single-feature LSTM-based classifiers with either log-mel spectrograms or modulation spectrograms as input features.}	 
	\label{tab:results_single}	
\end{table}

\begin{figure}[t]
	\centering
	\includegraphics[width=1.0\textwidth]{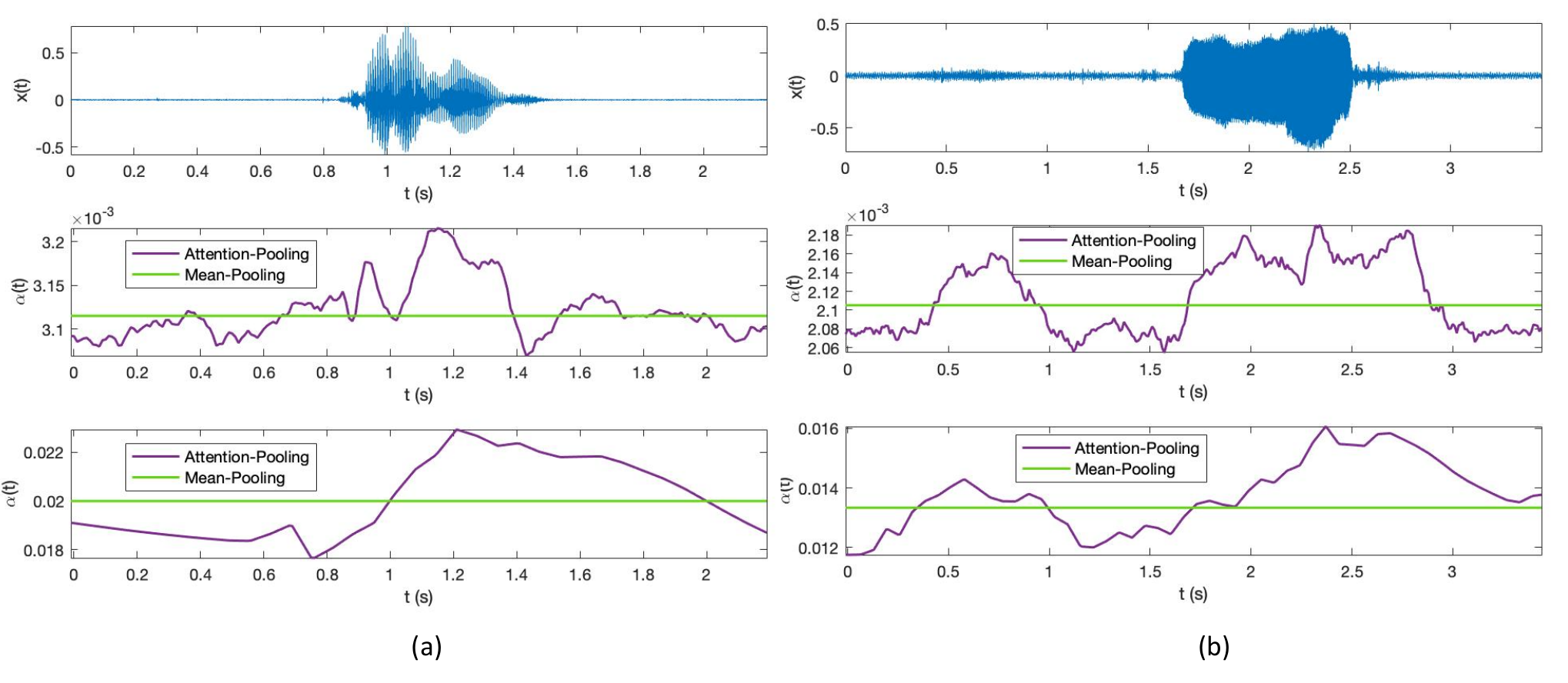}
	\caption{Mean-Pooling and Attention-Pooling weights for an utterance with (a) high intelligibility and (b) low intelligibility. Top: Waveform. Middle: Weights corresponding to the Mean-Pooling (green line) and Attention-Pooling approaches (violet line) for the LSTM-based system with log-mel spectrograms as input features.  Bottom: Weights corresponding to the Mean-Pooling (green line) and Attention-Pooling (violet line) approaches for the LSTM-based system with modulation spectrograms as input features. Both utterances correspond to the word ``jowls".}
	\label{fig:weights}
\end{figure}

As an example of the behavior of the attention mechanism, Figure \ref{fig:weights}(a) depicts the waveform (top) of an utterance with high intelligibility and the corresponding Mean-Pooling and Attention-Pooling weights when using log-mels as features (middle) and when using modulation spectrograms (bottom). Figure \ref{fig:weights}(b) displays the same information for a low intelligibility  case.  As a general remark, for all cases,  in contrast to the uniformity of Mean-Pooling weights, Attention-Pooling ones show a significant variation with time, depending on the relevance to the task that this mechanism assigns to each temporal frame. Attention weights for both kind of features exhibit similar trends although the weight curves corresponding to modulation spectrograms are smoother due to the larger analysis windows used for their computation in comparison to log-mels. From the analysis of these graphs, it can be observed that, although larger weights are assigned to high energy speech segments, low energy frames corresponding to pauses or speech artifacts also contribute to the final utterance representation as their weights are greater than zero. For example, the segment from $0.5$ to $0.8~s$ in the low intelligibility utterance corresponds to a hesitation and the corresponding attention weights present high values. This fact suggests the importance of frames related to disfluencies or rhythmic disturbances for the determination of the intelligibility level in both, log-mel and modulation domains.

\paragraph{Results of the fused systems}

Table \ref{tab:results_combined} contains the results attained by the systems where the two type of features, log-mel and modulation spectrograms, are combined. In particular, the corresponding classification rates for the two fusion strategies proposed in this work: fusion at decision level (\emph{Late Fusion}) and at utterance level (\emph{WP Fusion}), and the three WP methods under consideration: Basic, Mean-Pooling and Attention-Pooling, are reported. \textcolor{black}{It also contains the accuracies achieved by the combined SVM-based systems for both, late and early fusion. Together to these classification rates, Figure \ref{fig:results} shows the corresponding 95\% confidence intervals.}
	
\textcolor{black}{In the case of SVM, both fusion strategies outperform the corresponding single-feature systems, being \emph{late fusion} the method that produces best results. However, these accuracies are significantly worse than those attained by any of the individual and fused LSTM-based systems.}

\textcolor{black}{Analyzing the behaviour of the fused LSTM-based systems, it can be observed that} both combination approaches improve the performance of the individual systems for the three WP schemes. As in the case of the single-feature systems, \emph{LSTM Attention-Pooling} is the best option followed by \emph{LSTM Mean-Pooling}. These results corroborate our hypotheses that log-mel and modulation spectrograms carry complementary information about the level of intelligibility of an utterance and that the attention mechanism is an useful tool for this task, no matter the features used as input to the LSTM-based systems.

The comparison between the two fusion strategies \textcolor{black}{in LSTM-based systems} shows that the combination at utterance level is more beneficial than the mixture of the outputs of the respective single-feature systems. Specifically, the best system (\emph{WP Fusion} + \emph{Attention-Pooling}) achieves a relative error reduction of \textcolor{black}{
$39.41~\%$ with respect to \emph{SVM} + \emph{Late Fusion},} $14.84~\%$  with respect to \emph{Log-mel spectrograms} + \emph{Attention-Pooling},  of $19.01~\%$  with respect to \emph{Modulation spectrograms}  + \emph{Attention-Pooling} and of $9.97~\%$  with respect to the second best system (\emph{Late Fusion} + \emph{Attention-Pooling}). \textcolor{black}{In addition, the differences in performance between \emph{WP Fusion} + \emph{Attention-Pooling} and all the remaining systems are statistically significant, as can be observed in Figure \ref{fig:results}.}

\begin{table}[t]
	\centering	
	
	\begin{tabular}{|c|c|c|}
		\hline
		\textbf{System} & \textbf{Type of combination} & \textbf{Accuracy [\%]}  \\
		\hline
		\hline
		\textcolor{black}{SVM} & \textcolor{black}{Late Fusion} & \textcolor{black}{56.33~\%  $\pm$  0.79~\%} \\
		\hline
		\textcolor{black}{SVM} & \textcolor{black}{Early Fusion} & \textcolor{black}{50.10~\%  $\pm$  0.75~\%} \\
		\hline
		\hline
		Basic LSTM - No Pooling & Late Fusion & 63.47~\%  $\pm$  0.67~\% \\
		\hline
		LSTM Mean-Pooling & Late Fusion & 68.16~\%  $\pm$  0.61~\% \\
		\hline
		LSTM Attention-Pooling & Late Fusion & \textbf{70.61~}\%  $\pm$ \textbf{0.52~} \% \\
		\hline
		\hline
		Basic LSTM - No Pooling & WP Fusion & 64.60~\%  $\pm$  0.51~\% \\
		\hline
		LSTM Mean-Pooling & WP Fusion & 69.00~\%  $\pm$  0.55~\% \\
		\hline
		LSTM Attention-Pooling & WP Fusion & \textbf{73.54~\%}  $\pm$  \textbf{0.48~\%} \\
		\hline
	\end{tabular}	
	\caption{\textcolor{black}{Average classification rates [\%] and the corresponding standard deviations achieved by the SVM-based reference system with two types of combination strategies: late and early fusion,} and the LSTM-based classifiers with two types of combination strategies:  late fusion and WP fusion.}	 
	\label{tab:results_combined}	
\end{table}

\begin{figure}[t]
	\centering
	\includegraphics[width=1.0\textwidth]{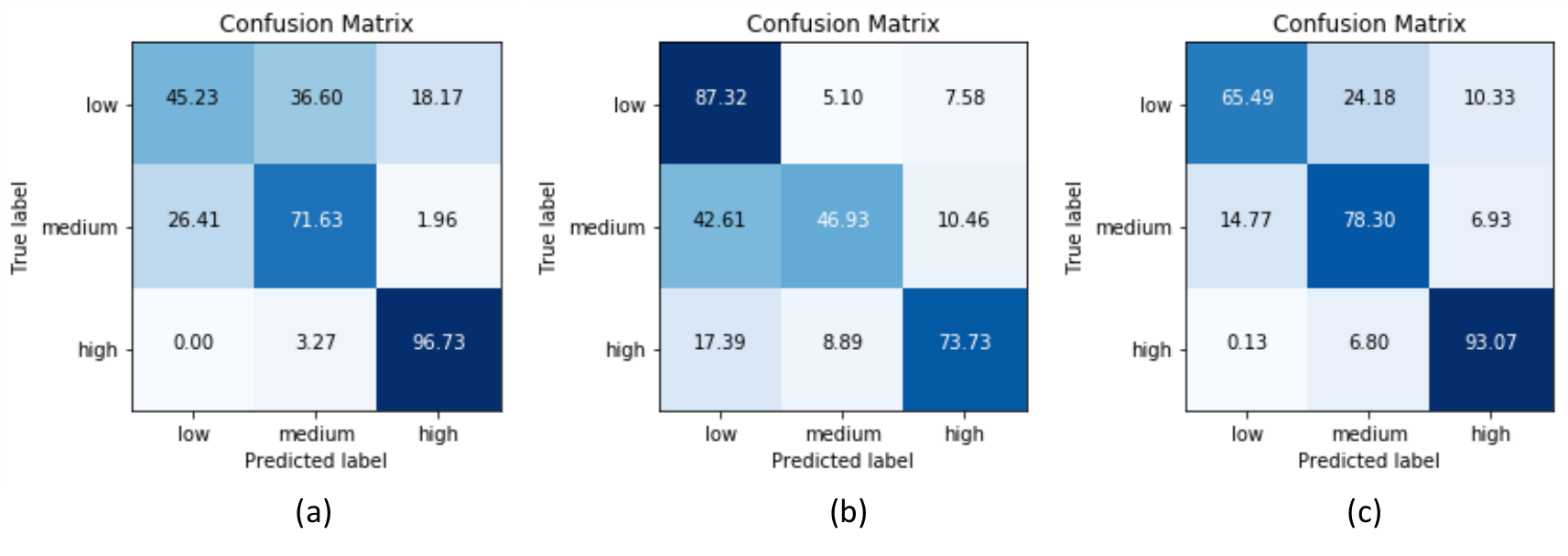}
	\caption{\textcolor{black}{Confusion matrices [\%] for the LSTM-based systems with attention and three different input features: (a) Log-mel spectrograms; (b) Modulation spectrograms; (c) Combined system with WP fusion.}}
	\label{fig:confusion_matrices}
\end{figure}

\textcolor{black}{In order to perform a more detailed analysis about the complementarity of log-mel and modulation spectrograms, we have obtained the confusion matrices produced by the single-feature systems \emph{Log-mel spectrograms} + \emph{Attention-Pooling} and \emph{Modulation spectrograms}  + \emph{Attention-Pooling}, and the fused system \emph{WP Fusion} + \emph{Attention-Pooling}. These confusion matrices are represented in, respectively, Figure  \ref{fig:confusion_matrices} (a), (b) and (c). In the three graphics, the rows correspond to the correct class, the columns to the hypothesized one and the values in them are computed as the average over the test utterances belonging to the fourth fold.}

\textcolor{black}{As can be observed, in the case of log-mel spectrograms, the less confusable class (with a classification rate greater) is high intelligibility, whereas there is about $37~\%$ of low intelligibility utterances that are misclassified as medium ones. For modulation spectrograms, the class presenting the best performance is low intelligibility. In this case, more than $40~\%$ of medium intelligibility items are incorrectly assigned to the low class. This distinctive behaviour supports the hypothesis that both sets of features carry complementary information, so their combination can outperform the individual systems, as is corroborated by the results in Table \ref{tab:results_combined} and Figure \ref{fig:results}, and the confusion matrix of the fused system represented in Figure \ref{fig:confusion_matrices}.}

\subsection{\textcolor{black}{Computation complexity}}
\label{subsec:computational_complexity}

\textcolor{black}{In this Subsection, the computational complexity of the LSTM-based systems developed in this work is analyzed. Firstly, for all WP schemes, the sequence learning performed by the LSTMs is the process that has more impact on the overall computational cost. For that, we ignore the effect of the remaining stages when estimating the complexity of the whole systems.}

\textcolor{black}{According to \cite{Hochreiter1997}, the LSTM algorithm is very efficient, exhibiting  a time complexity per time step of $O(W)$, where $W$ is the number of trainable parameters (weights).  Therefore, the overall complexity of a single LSTM-based system is $O(W \times L)$, where $L$ is the LSTM input sequence length. The complexity of the combined systems can be approximated as the sum of the complexity of the corresponding individual systems.}

\textcolor{black}{Table \ref{tab:computational_oomplexity} contains the values of $W$, $L$ and $O(.)$ for the LSTM-based systems with log-mel and modulation spectrograms as input features and the fused system. It also contains the relative complexity with respect to the system that uses log-mel spectrograms as input features. As can be observed,  when utilizing modulation spectrograms, the complexity is four times less than for log-mels, and, therefore it is a good alternative in cases where there are computational constraints. Otherwise, the complexity of the combined system is about $ 1.27$ times that of log-mels.  This increase in complexity is acceptable, taking into account the benefits achieved in terms of accuracy.}

\begin{table}[t]
	\centering	
	
	\begin{tabular}{|c|c|c|c|c|}
		\hline
		\textcolor{black}{\textbf{System}} & \textcolor{black}{$\mathbf{W}$} & \textcolor{black}{$\mathbf{L}$}  & \textcolor{black}{$O(.)$} & \textcolor{black}{\textbf{Relative complexity}} \\
		\hline
		\hline
		\textcolor{black}{Log-mel spectrogram} & \textcolor{black}{24832} & \textcolor{black}{700}  & \textcolor{black}{16.58~M} &  \textcolor{black}{1} \\
		\hline
		\textcolor{black}{Modulation spectrogram} & \textcolor{black}{42240} & \textcolor{black}{110} & \textcolor{black}{4.43~M} & \textcolor{black}{$\approx$ 0.27} \\
		\hline
		\textcolor{black}{Combination} & \textcolor{black}{67072} &  \textcolor{black}{|}  & \textcolor{black}{21.01~M} & \textcolor{black}{$\approx$ 1.27} \\
		\hline
	\end{tabular}	
	\caption{\textcolor{black}{Computational complexity of LSTM systems based on log-mel spectrograms, modulation spectrograms and their fusion.}}	 
	\label{tab:computational_oomplexity}	
\end{table}

\subsection{\textcolor{black}{Limitations of the study}}
\label{subsec:.limitations_study}

\textcolor{black}{There are two major limitations in this work that could be addressed in future research. The first is the lack of diversity of the training data, in the sense that the number of speakers (15 speakers) and intelligibility scores (15 different scores in the range from 0 to 100) is scarce. In addition, the data is imbalanced with respect to the genre of the speakers (11 men vs. 4 women). Due to the difficulty of collecting data for this task, the exploration of the appropriate data augmentation methods could be an interesting option for overcoming, at least partially, this issue.}
	
\textcolor{black}{The second limitation concerns the accuracy achieved by the best of our systems, that implies that around 25\% of the audio files are incorrectly classified. On the one hand, classification rates could be improved by means of the use of the aforementioned data augmentation techniques. On the other hand, it must be taken into account that speech intelligibility classification is a challenging task, not only for machines but also for humans, and depends to some extent on the intrinsic characteristics of the utterance to be listened. In this sense, the audio recordings in the database contain utterances of very different types: digits, radio alphabet, computer commands, simple short words and multisyllable complex words. The characteristics of these broad groups of utterances are rather different regarding, among others, their duration, pronunciation difficulty, and phoneme similarity and confusability. An exhaustive analysis of the classification errors as a function of the aforementioned factors could give insight into the design of the best set of words to be used for intelligibility measurement. }

\section{Conclusions and future work}
\label{sec:conclusions} 

In this paper, we have extended our previous work about the development of an automatic non-intrusive system for speech intelligibility level classification based on attention LSTM networks. We present  two main contributions. In the first one, we have proposed the use as input features of per-frame modulation spectrograms, instead of compact representations derived from them that discard important temporal information. As modulation spectrograms capture long-term phenomena typically presented in pathological speech, whereas log-mel spectrograms are more related to short-term events, in the second contribution we have explored two different strategies for the combination of both kind of \textcolor{black}{per-frame} features \textcolor{black}{into the LSTM-based architecture}: at decision level (late fusion) and at utterance level (WP fusion). In both cases, three different weighting schemes in the LSTM architecture have been evaluated: basic LSTM, LSTM with Mean-Pooling and LSTM with Attention-Pooling.

The developed systems have been assessed over the UA-Speech database that contains dysarthric speech with different levels of severity. \textcolor{black}{First, it can be observed that LSTM-based systems significantly outperform the performance of traditional SVM-based systems. Second},  results have shown that attention LSTM networks are able to adequately modeling the modulation spectrograms sequences producing similar classification rates as in the case of the \textcolor{black}{LSTM-based system with} log-mel spectrograms. In all cases, the weighting scheme based on the attentional mechanism is the one that achieves the best results, so it is clear that learning the contribution of each frame to the task improves the system performance. \textcolor{black}{Third}, both combination strategies, late and WP fusion, outperform the single-feature systems, suggesting that log-mel and modulation spectrograms carry complementary information than can be effectively exploited by the LSTM-based architectures. Best results have been achieved by the combined system with WP fusion and Attention-Pooling, obtaining relative error reductions \textcolor{black}{of 39.41 \% with respect to SVM + Late Fusion, and}  of $14.84~\%$  and $19.01~\%$ with respect to log-mel spectrograms and Attention-Pooling and modulation spectrograms and Attention-Pooling, respectively. \textcolor{black}{All these performance differences are statistically significant.}

\textcolor{black}{Regarding the applicability of this study, the assessment of the speech intelligibility level can be useful, among others, as part of diagnosis aid systems for certain diseases or the monitoring of patients following logopedic or other medical therapies. This is, for example, the case of PD where the Unified Parkinson's Disease Rating Scale (UPDRS), which is commonly used to track the progression of this illness, contains two assessment criteria related to the speech communication skills of the patient \cite{Goetz2003}.  The gold standard for judging the comprehensibility level of a patient's speech consists of carrying out several tests where the subject pronounces various words and/or sentences. Then, one or more specialists listen to these utterances and assign the intelligibility scores according to the percentage of words they understood. Nevertheless, multiple benefits can be achieved from the automatization of this task, as is proposed in this work. Firstly, it allows more time to the medical staff to carry out other activities. Secondly, the specialists' criterion to assess the intelligibility is in a way subjective as it relies on their own hearing skills that might be influenced by their familiarization with pathological speech \cite{Landa2014}. Finally, it provides an objective and reproducible measure.}

For future work, we plan to study the use of weights derived from auditory saliency features \cite{Kaya2017} in the attention mechanism, following the promising results achieved in \cite{Gallardo-Antolin2019a} for cognitive load estimation from speech. In addition, we plan to extend our research towards the exploration of the more suitable data augmentation techniques for improving the training process and performance of the speech intelligibility level classification system. \textcolor{black}{Finally,  the analysis of the classification errors as a function of several factors, such as utterance duration, pronunciation difficulty and phoneme confusability could help to select the most appropriate set of words for intelligibility measurement.}

\section*{Acknowledgments}

The work leading to these results has been partly supported by the Spanish Government-MinECo under Projects TEC2017-84395-P and TEC2017- 84593-C2-1-R. The authors wish to acknowledge Dr. Mark Hasegawa-Johnson for making the UA-Speech database available.


\bibliography{mybibfile_nc}

\end{document}